\documentclass{emulateapj}
\shorttitle{The systematic and random errors in the PPMXL catalog}
\shortauthors{Wu et al.}
\begin{document}
\title{Independent determination of the systematic and random errors of the proper motions in the PPMXL catalog using quasars}
\author{Zhen-Yu Wu, Jun Ma, Xu Zhou}
\affil{Key Laboratory of Optical Astronomy, National Astronomical Observatories, Chinese Academy of Sciences, 20A Datun Road, Beijing 100012, China}
\email{zywu@nao.cas.cn}
\begin{abstract}
Using a sample of 117053 quasars identified in the PPMXL catalog, the systematic errors of the proper motions in the PPMXL catalog are estimated independently as $-2.0$ and $-2.1$ mas yr$^{-1}$ in  $\mu_{\alpha}\cos\delta$ and $\mu_{\delta}$, respectively. For objects with SDSS $r$ magnitude  between 16.0 and 21.0 mag and SDSS $g-r$ color between $-0.4$ and 1.2, there are no obvious magnitude and color dependence in the systematic errors of the proper motions in the PPMXL catalog. The random errors of the proper motions in the PPMXL catalog change from $\sim4.0$ to $\sim8.0$ mas yr$^{-1}$ along with the increase of the magnitude of the objects. No color dependence in the random errors of the proper motions in the PPMXL catalog is found. There are obvious right ascension $\alpha$ dependence of  the systematic errors in both components of the proper motions in the PPMXL catalog. For quasars with declination $\delta>0$,  the systematic errors of $\mu_{\delta}$ in the PPMXL catalog increase linearly with $\delta$ with a slope of 0.07 mas yr$^{-1}$ deg$^{-1}$. 

For comparison, using a subsample of quasars identified both in the PPMXL and SDSS DR7 catalogs,  the systematic errors of the USNO-SDSS proper motions in the SDSS DR7 catalog are derived as  $0.9$ and $0.5$ mas yr$^{-1}$ in  $\mu_{\alpha}\cos\delta$ and $\mu_{\delta}$, respectively. There are no obvious magnitude, color, and $\alpha$ dependence in the systematic errors of the USNO-SDSS proper motions. For quasars with declination $\delta>0$,  the systematic errors of $\mu_{\delta}$ in the SDSS DR7 catalog increase linearly with $\delta$ with a slope of 0.03. The random errors of the USNO-SDSS proper motions change from $\sim2.5$ to $\sim6.0$ mas yr$^{-1}$ along with the change of the magnitude and color of the objects. Combining the 2MASS or SDSS observation data can significantly reduce the systematic errors of the proper motions in the initial photographic catalog. 
\end{abstract}
\keywords{Stars --- Quasars and Active Galactic Nuclei --- Data Analysis and Techniques}

\section{Introduction}
Absolute proper motions of stars are very important data for the kinematic studies of the Galaxy. For the last two decades, great efforts have been made to acquire high precise and accurate absolute proper motions of stars covering the entire sky. The Hipparcos catalog is considered as the realization of the International Celestial Reference System (ICRS) at optical band. The systematic error of the absolute proper motions in the Hipparcos catalog with respect to ICRS is estimated to be $0.25$ mas yr$^{-1}$. The Hipparcos catalog contains about 0.12 million stars and is complete down to $V=7.3 - 9.0$ mag \citep{es97}. 

The Tycho-2 catalog \citep{ho00} is the first densification of the Hipparcos catalog. The proper motions in the Tycho-2 catalog are on the system of the Hipparcos with a deviation under 0.5 mas yr$^{-1}$ \citep{ur00}. The Tycho-2 catalog contains 2.5 million stars and is complete down to $V\sim11.5$ mag. The ASCC-2.5 catalog is based on large, modern, and high-precision catalogs of the Hipparcos-Tycho family, including the Tycho-2 catalog, and provides the most complete all-sky catalog of 2.5 million stars having uniform high-precision astrometric and photometric data \citep{kh01}. 

The UCAC2 catalog presents its proper motions on the Hipparcos system with the nominal errors of 1 to 3 mas yr$^{-1}$ for stars up to $V \sim 12$ mag and about 4 to 7 mas yr$^{-1}$ for fainter stars up to $V \sim 16$ mag. The systematic errors of the proper motions in the UCAC2 catalog are in the range from 0.5 to 1.0 mas yr$^{-1}$ \citep{za04}. The UCAC2 catalog contains 48 million objects with limiting magnitude of $R\sim 16$ mag. But UCAC2 catalog is not a all-sky catalog, most of data in the regions with declination $\delta>+40\degr$ do not exist in this catalog.

\citet{ro08} presented the PPMX catalog which including positions and absolute proper motions of $\sim18$ million stars with limiting magnitude $\sim 15$ mag in a red band. Very recently, \citet{za10} published the all-sky UCAC3 catalog which containing over 100 million objects with slightly deeper limiting magnitude than that of the UCAC2 catalog.

The absolute proper motions in the above-mentioned catalogs can only be used in the kinematic studies of stars in the neighborhood of the Sun due to their brighter limiting magnitudes of stars in those catalogs \citep[e.g.][]{wu09}. In recent decade, a few of catalogs are available for stars with more fainter limiting magnitudes. \citet{ha01} presented the SuperCOSMOS catalog whose zero-point of the absolute proper motion system is fixed to the extragalactic reference frame. The mean error of the proper motions in this catalog changes from typically 10 mas yr$^{-1}$ for objects with $R\sim 18$ mag to 50 mas yr$^{-1}$ for objects with $R\sim 21$ mag.  The zero-point error in the proper motions of the SuperCOSMOS catalog is less than 1 mas yr$^{-1}$ for objects with $R > 17$ mag, and is not larger than 10 mas yr$^{-1}$ for objects with $R < 17$ mag. The GSC\,II catalog presents astrometry, photometry, and classification for about 946 million objects, but the proper motion data are not released to the scientific community due to the significant systematic errors in the derived proper motions for objects in the southern hemisphere \citep{la08}.

The USNO-B catalog presents proper motions for about one billion objects based on data obtained from scans of 7435 Schmidt plates \citep{mo03}. But the proper motions in this catalog are relative, not absolute, which limits their use in the kinematic studies of stars in the Galaxy \citep{wu11}. Combining the USNO-B and Sloan Digital Sky Survey (SDSS) catalogs, \citet[hereafter USNO-SDSS proper motion]{mu04,mu08} presented an improved absolute proper-motion catalog . The USNO-B positions are recalibrated using the galaxies identified in the SDSS catalog, and the proper motions are placed on an absolute reference frame and are recomputed using both the USNO-B and SDSS positions. The random proper-motion error is only $\sim 3$ mas yr$^{-1}$ for object with $r < 18$ mag and $\sim 5$ mas yr$^{-1}$ for object with $r < 20$ mag. The systematic error is typically an order of magnitude smaller \citep{bo10}. But the available USNO-SDSS proper motion data are limited by the regions covered by the SDSS observation.  

\citet{fe10} presented the XPM-1.0 catalog which including absolute proper motions of about 280 millions stars covering the entire sky. The positions in the USNO-A2.0 \citep{mo98}  and 2MASS catalogs were used to calibrate the proper motions of XPM-1.0. The zero-point of the absolute proper motions in the XPM-1.0 catalog was linked to more than one million galaxies. The random error of the proper motions in this catalog varies  from 3 to 10 mas yr$^{-1}$ depending on magnitudes. The systematic error of the proper motions in the Northern hemisphere is $\sim 0.3$ mas yr$^{-1}$, and is $\sim 1.0$ mas yr$^{-1}$ in the Southern hemisphere \citep{fe10}. But, the faint end of XPM-1.0 catalog is limited by the limiting magnitude of the 2MASS catalog.

Very recently, the PPMXL catalog was released \citep{ro10}. Combing USNO-B catalog with 2MASS \citep{sk06} catalog, the new mean positions and absolute proper motions of objects listed in the USNO-B catalog were derived with least-squares adjustment and reduced to the ICRS system using the PPMX catalog \citep{ro08}. The PPMXL catalog contains $\sim910$ millions objects, of which, 410 millions objects have photometric data from 2MASS catalog. At its bright end, the PPMXL catalog is merged with the PPMX catalog. So, PPMXL is the largest collection of absolute proper motions on the ICRS system and is complete from the brightest stars down to $V\sim 20$ mag. The mean errors of the proper motions in PPMXL catalog change from 4 mas yr$^{-1}$ to more than 10 mas yr$^{-1}$ depending on the magnitudes.

As the largest available absolute proper motion catalog with very fainter limiting magnitude, the PPMXL catalog is a very important resource in the kinematic studies of the Galaxy. Before using the proper motion data in the PPMXL catalog, it is very important to understand the systematic and random errors of the proper motions in this catalog. \citet{ro10} have corrected the plate-dependent distortions of the proper motions during the construction  procedure of  the PPMXL catalog. \citet{ro10} have pointed out that the magnitude- and color-dependent systematic errors in the PPMXL catalog are difficult to be determined due to no independent reference on ICRS at fainter magnitude. On the other hand, it is also difficult to distinguish between systematic errors and different kinematics for different populations in the Galaxy.

Quasars are very distant extragalactic objects and their proper motions can be considered as zero. On the other hand, the PSFs of quasars in an image are close to those of the stars, the systematic deviation from zero within the proper motions of the quasars and their standard deviation represent the systematic and random errors in the proper motions of the stars. In this work, a largest available quasars sample is used to analyze the systematic and random errors in the proper motions of the PPMXL catalog.

We present the construction  procedure of  the PPMXL catalog in Section 2. The quasar sample identified in the PPMXL catalog is presented in Section 3. In section 4, the systematic and random errors in the proper motions of the PPMXL catalog are derived using the quasar sample. For comparison, the systematic and random errors in the USNO-SDSS proper motions are derived in Section 5. Our discussion and conclusions are given in Section 6.

\section{Construction of the PPMXL catalog}
First, the USNO-B catalog was cross-identified with the PPMX catalog. The common stars in the PPMX catalog fainter than Tycho-2 limits were used, for each field (corresponding to a Schmidt plate), to derive the mean deviations in right ascension $\Delta \alpha$ and declination $\Delta \delta$ at the field epoch from the USNO-B catalog. This step corrected the positions in the USNO-B catalog to the ICRS. Then, the field-corrected USNO-B catalog was cross-identified with the 2MASS catalog. For each common star, its positions at different epochs in the USNO-B catalog and at the observational epoch of the 2MASS catalog were recalculated with a least-squares method. In the least-squares adjustment procedure, individual weight  for each star at different epoch is attributed. After the least-squares adjustment, a preliminary catalog with the new position and proper motion for each common star was constructed.

To link the proper motion system of the preliminary catalog to the ICRS, stars in the preliminary catalog with $K_{s}$ between 12 and 13 mag were chosen to represent the preliminary system which called PS1, and the PPMX catalog was chosen as the optical representation of the ICRS. Before linking to the PPMX catalog, the obvious plate-dependent distortions of the proper motions of the PS1 in the south of $-20\degr$ declination were corrected using the proper motions of the UCAC3 catalog. The proper motions of the PS1 are left unchanged north of $-20\degr$ declination. The UCAC3-corrected PS1 system is called PS2. Then, using the proper motions in the PPMX catalog, the PS2 was corrected to ICRS, the resulting system is called PS3.

Above-mentioned procedure only considered stars having 2MASS observations. To link the proper motions of objects fainter than the 2MASS limits on the ICRS, the least-squares adjustment were recalculated giving weight zero to the 2MASS observations. For stars in PS3, the differences in the proper motions with and without including 2MASS observations were derived. Then, these differences were used to corrected the proper motions of objects without including 2MASS observations to the ICRS represented by the PPMX catalog. 

So far, the proper motions of all objects in the PPMXL catalog are on the ICRS represented by the PPMX catalog. At the final stage, PPMXL is merged with PPMX. For an object in the PPMX catalog, if there is no object matched in the PPMXL or the error of its proper motions in the PPMX is smaller than that in the PPMXL, the data in the PPMX will be added to the PPMXL or replace that in the PPMXL.

\section{Quasar sample}
The 13th edition of the catalog of quasars and active nuclei \citep[hereafter VV13]{ve10} is used as the primary source for known quasars. This catalog contains 133336 quasars which includes all known quasars when this catalog was compiled. Two parameters of quasars: magnitude and coordinate can be used to cross-identify the common object between the VV13 and the PPMXL catalogs. Because the photometric data in these two catalogs are photographic magnitude with very large uncertainties, only the coordinates of quasars in these two catalogs were used. For a quasar listed in the VV13 catalog, if only one object is matched in the PPMXL catalog within a given search radius, then this object is considered as the same one in both two catalogs. If the search radius is too small, there is no any matched object between these two catalogs. When the search radius increases, more objects will be found within the given search radius, but no additional data can be used to distinguish the true matched one. Using a similar method of \citet{so09} used in their construction of the large quasar astrometric catalog (LQAC), different search radii were used to find the one and only matched quasars between the VV13 and the PPMXL catalogs.

\begin{figure*}
\begin{center}
\includegraphics[width=160mm,height=120mm]{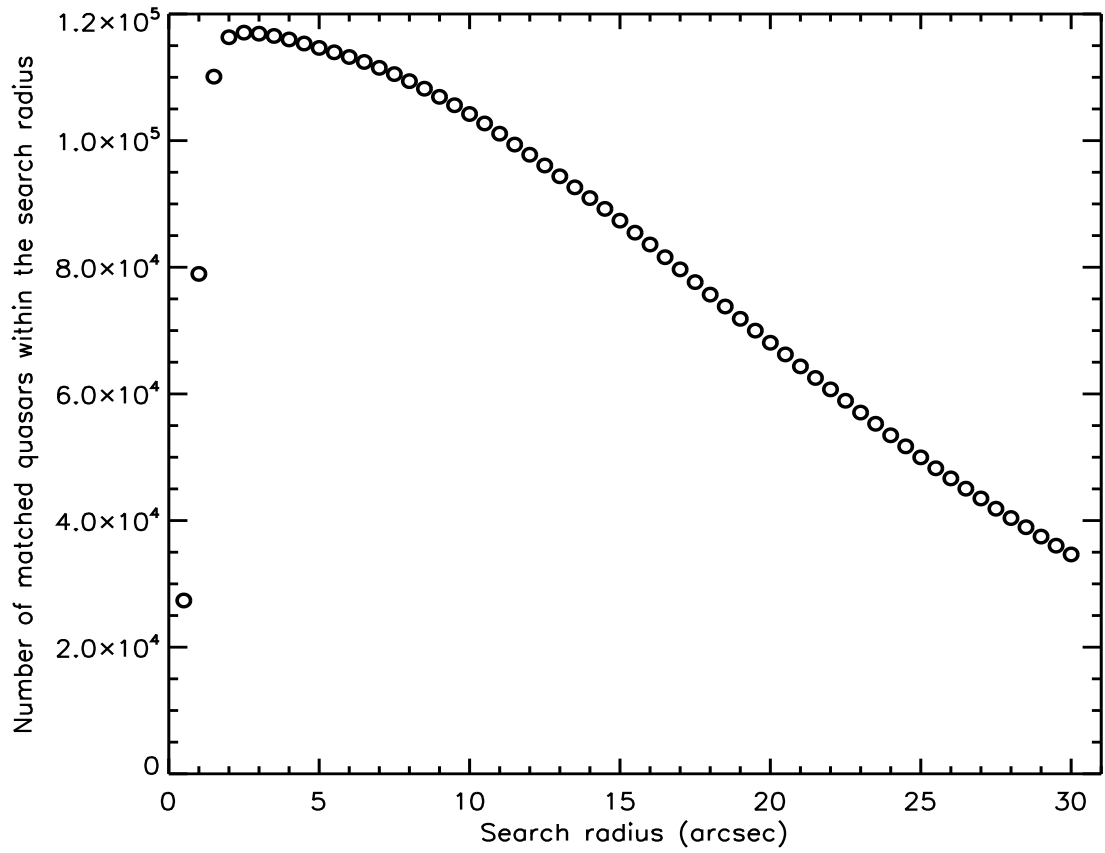}
\caption{Number of the uniquely cross-identified quasars between the VV13 and the PPMXL catalogs within the given search radius. A search radius of 2.5 arcsec was adopted finally.\label{fg1}}
\end{center}
\end{figure*}

\begin{figure*}
\begin{center}
\includegraphics[width=160mm,height=80mm]{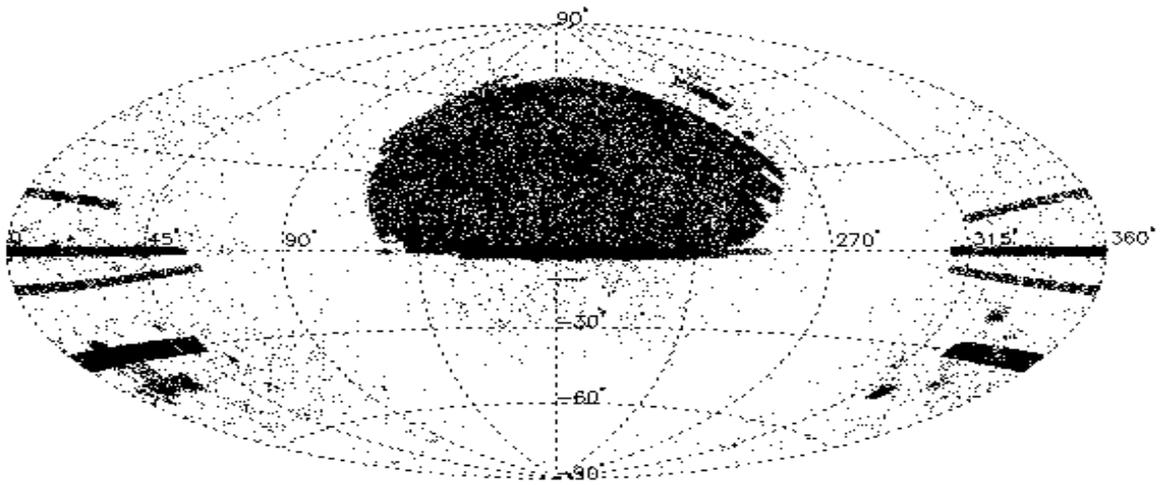}
\caption{Spatial distribution of 117053 quasars in the PPMXL catalogs in a J2000 equatorial reference frame with a Aitoff projection.\label{fg2}}
\end{center}
\end{figure*}

Figure \ref{fg1} shows the number of matched quasars between the VV13 and the PPMXL catalogs within different search radii from 0.5 to 30 arcsec with a step of 0.5 arcsec. It can be seen from Figure \ref{fg1} that the number of matched quasars increases very quickly to its maximum from 0.5 arcsec to 2.5 arcsec, then this number decreases slowly along with the increased search radius. Using the search radius of 2.5 arcsec, 117053 quasars are matched and are taken as the members of our quasar sample. For the matched quasars, their data in the PPMXL catalog were extracted using the catalog access tool VizieR. Figure \ref{fg2} shows the spatial distribution of the 117053 quasars in our sample. The distribution of quasars is very non-uniform, most of quasars are located at the limited regions and are spectroscopically confirmed with the observations of Sloan Digital Sky Survey (SDSS) \citep{ab09,sc10}. With a search radius of 1 arcsec and using the catalog access tool VizieR, the coordinates of quasars listed in the PPMXL catalog were used to cross-identified with the seventh data release (DR7) of SDSS \citep{ab09}, 97150 quasars in our sample were identified in the SDSS DR7 catalog.

\begin{figure*}
\begin{center}
\includegraphics[width=160mm,height=70mm]{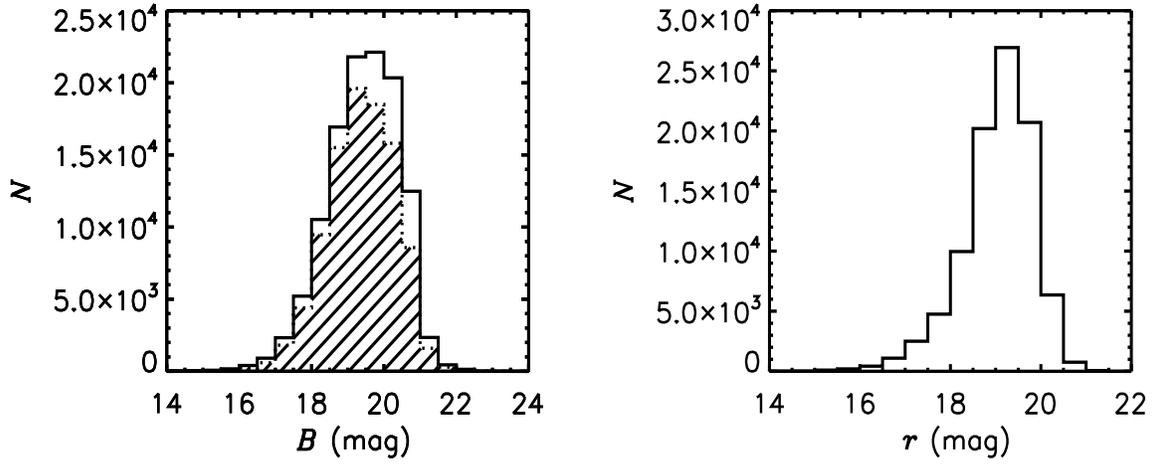}
\caption{Magnitude distributions of quasars in the PPMXL catalog. Left panel: photographic $B$ magnitude distribution. The shadow area is the distribution of cross-identified quasars between the PPMXL and the SDSS DR7 catalogs. Right panel: SDSS $r$ magnitude distribution.\label{fg3}}
\end{center}
\end{figure*}

\begin{figure*}
\begin{center}
\includegraphics[width=160mm,height=70mm]{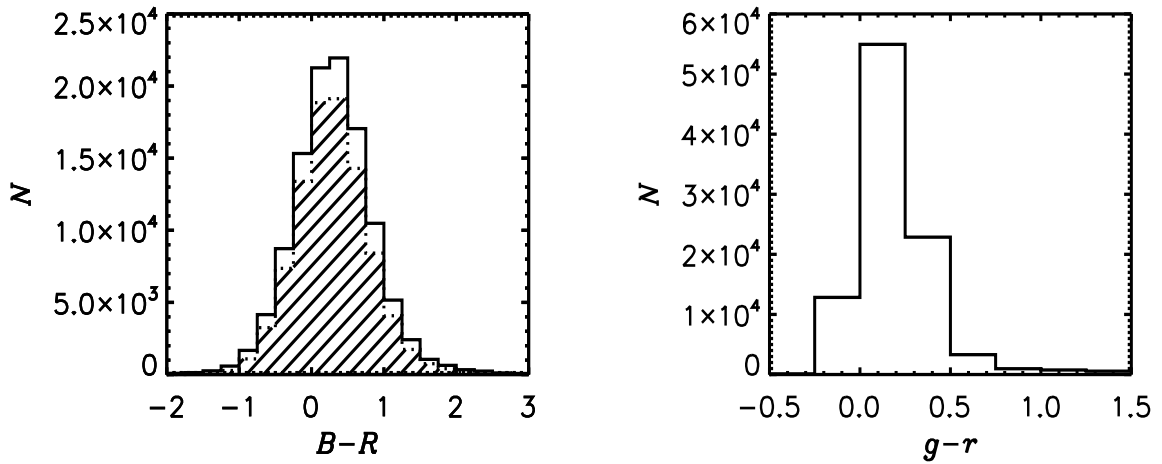}
\caption{Color distributions of quasars  in the PPMXL catalog. Left panel: photographic $B-R$ color distribution. The shadow area is the distribution of cross-identified quasars between the PPMXL and the SDSS DR7 catalogs. Right panel: SDSS $g-r$ color distribution.\label{fg4}}
\end{center}
\end{figure*}

Figures \ref{fg3} and \ref{fg4} show the magnitude and color distribution of quasars in our sample, respectively. The $B$ and $R$ magnitudes used in Figures \ref{fg3} and \ref{fg4} are photographic magnitudes derived from the PPMXL catalog. The shadow regions in the left panels of Figures \ref{fg3} and \ref{fg4} represent the distribution of the quasars  whose SDSS data are available. The SDSS $r$ magnitude and $g-r$ color distributions for the quasars in our sample are plotted in the right panels of Figures \ref{fg3} and \ref{fg4}, respectively. Figure \ref{fg3} shows that most of quasars in our sample have the photographic $B$ magnitudes between 17.0 and 21.5 mag and have SDSS $r$ magnitudes between 17.0 and 20.5 mag. Figure \ref{fg3} indicates that our present quasar sample represents the objects with fainter magnitudes in the PPMXL catalog. In our quasar sample, only  13520 quasars have the observation data from the 2MASS catalog. Thus, most of the proper motions of quasars in our sample were derived only from the photographic data. Figure \ref{fg4} shows that most of quasars in our sample are within the range of $B-R$ between $-1.0$ and $2.0$ and of SDSS $g-r$ between $-0.2$ and $0.8$. Figure \ref{fg4} also indicates that our present quasar sample only represents the objects in the PPMXL catalog within the limited color range.

\section{The systematic and random errors of the proper motions in the PPMXL catalog}
We have assumed that a quasar's distance is very far that its proper motion cannot be detected by present astrometric observation in the optical band and can be considered as zero. Thus, for a sample of quasars, any deviation from zero of the mean of the proper motions of the sample can be considered as the systematic error of the proper motions, and the dispersion of the proper motions from the mean can be considered as the random error of the proper motions of the sample. Figure \ref{fg5} shows the histograms of the proper motions of the quasars in our sample in right ascension $\mu_{\alpha}\cos\delta$ and in declination $\mu_{\delta}$ with a bin of 1.0 mas yr$^{-1}$. Figure \ref{fg5} only shows the distributions of the proper motions for quasars whose proper motions with the absolute value less than 35 mas yr$^{-1}$ in each component. A Gaussian function was used to fit the distribution of the proper motions in each component of all quasars in our sample and was plotted as red solid line in each panel of Figure \ref{fg5}. The region under the red solid line in each panel includes proper motion data within the $3\sigma$ of the mean. The best-fitting parameters of the Gaussian function: the mean and the dispersion of the proper motions in each component are labeled in each panel of Figure \ref{fg5}. Figure \ref{fg5} indicates that the mean or the systematic error of the proper motions of quasars in the PPMXL catalog is $-2.0$ and $-2.1$ mas yr$^{-1}$ in $\mu_{\alpha}\cos\delta$ and $\mu_{\delta}$, respectively. The random error  of the proper motions of quasars in the PPMXL catalog is $5.3$ and $4.8$ mas yr$^{-1}$ in $\mu_{\alpha}\cos\delta$ and $\mu_{\delta}$, respectively. 

\subsection{The magnitude dependence of the systematic and random errors of the proper motions}
\begin{figure*}
\begin{center}
\includegraphics[width=160mm,height=140mm]{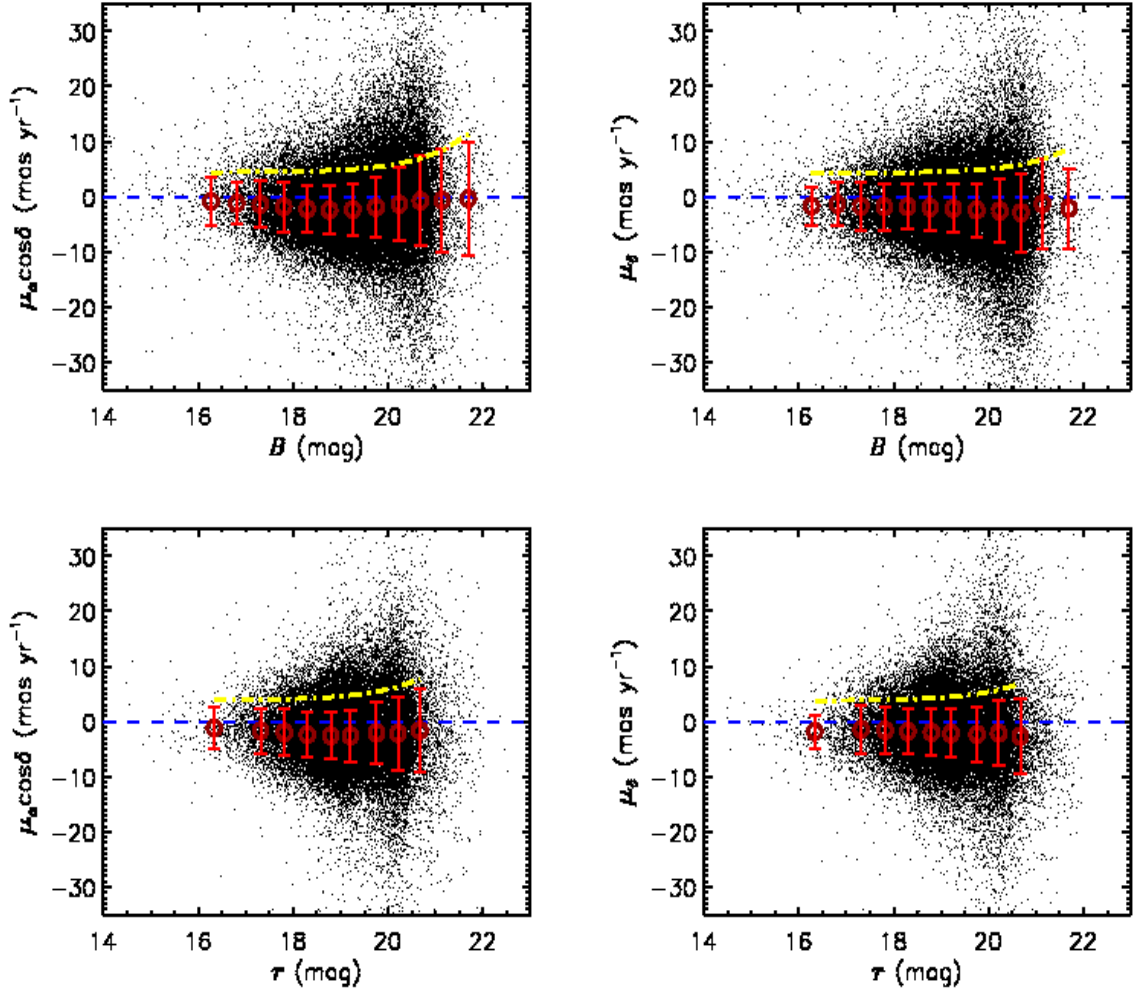}
\caption{Magnitude dependence of the proper motions of quasars in the PPMXL catalog. Top panels: photographic $B$ magnitude. Bottom panels: SDSS $r$ magnitude. The blue dashed line in each panel shows the proper motions with value of zero. The yellow dash dot line in each panel is the best-fitting function which describes the magnitude dependence of the random errors of proper motions in the PPMXL catalog: $\sigma_{\mu}=a+b\times\exp^{(m-c)}$, where $\sigma_{\mu}$ is the random error of proper motions in one component, $a$, $b$, and $c$ are the unknown parameters to be fitted, and $m$ is the $B$ or $r$ magnitude. Not all of the quasars in our sample are showed, but the proper motions in each component for all quasars in each magnitude bin are used to derive the systematic and random errors in that magnitude bin, which are showed as red open circles and error bars in each panel.\label{fg6}}
\end{center}
\end{figure*}

Figure \ref{fg6} shows the magnitude dependence of the proper motions of quasars in our sample in each component. The top two panels are based on the photographic $B$ magnitude,  and the bottom two panels are based on the SDSS $r$ magnitude. The blue dashed line in each panel shows the proper motions with value of zero. In order to derive the dependence of the proper motions in each component on the magnitude, quasars are divided into different magnitude groups with a bin of 0.5 mag. For photographic $B$ magnitude, quasars with $B$ magnitude in the range between 16.0 and 22.0 mag were used.  For SDSS $r$ magnitude, quasars with $r$ magnitude in the range between 16.0 and 21.0 mag were used. For each magnitude group of quasars, a Gaussian function was used to fit the proper motions in each component in that group, the mean and dispersion of the best-fitting Gaussian function corresponding to the systematic and random errors of proper motions of quasars in that group are showed as red open circle and error bar in each panel of Figure \ref{fg6}.

The left panels of Figure \ref{fg6} indicate that the systematic errors of $\mu_{\alpha}\cos\delta$ have the similar dependence on the  photographic $B$ magnitude and on the SDSS $r$ magnitude. The absolute value of the systematic deviation of $\mu_{\alpha}\cos\delta$ increases from the brightest end of the quasar magnitude group to its maximum, then decreases progressively to the faintest end of the quasar magnitude group. The rms of the systematic errors of $\mu_{\alpha}\cos\delta$ for different quasar magnitude groups is $0.7$ mas yr$^{-1}$ for photographic $B$ magnitude and $0.5$ mas yr$^{-1}$ for SDSS $r$ magnitude, respectively.  But this rms is smaller than the absolute systematic deviation of $2.0$ mas yr$^{-1}$ in $\mu_{\alpha}\cos\delta$ for the whole quasar sample by a factor of 3 -- 4.

Except the last two points in the faint photographic $B$ magnitude bins, the right panels of Figure \ref{fg6} indicate that the systematic errors of $\mu_{\delta}$ have the similar dependence on the  photographic $B$ magnitude and on the SDSS $r$ magnitude. The absolute value of the systematic deviation of $\mu_{\delta}$ decreases progressively from the brightest end of the quasar magnitude group to the fainter end of the quasar magnitude group. The rms of the systematic errors of $\mu_{\delta}$ for different quasar magnitude groups is $0.5$ mas yr$^{-1}$ for photographic $B$ magnitude and $0.4$ mas yr$^{-1}$ for SDSS $r$ magnitude, respectively. This rms is smaller than the absolute systematic deviation of $2.1$ mas yr$^{-1}$ in $\mu_{\delta}$ for the whole quasar sample by a factor of 4.

The red error bar in each panel of Figure \ref{fg6} shows the random error of the proper motions in each component in each quasar magnitude group. All of the four panels of Figure \ref{fg6} indicate that the random errors of proper motions in each component increase along with the magnitude of quasars.  The magnitude dependence of the random error of proper motions in each component can be described by a function: $\sigma_{\mu}=a+b\times\exp^{(m-c)}$, where $\sigma_{\mu}$ is the random error of proper motions in one component, $a$, $b$, and $c$ are the unknown parameters to be fitted, and $m$ is the $B$ or $r$ magnitude. In each panel of Figure \ref{fg6}, the yellow dash dot line shows the best-fitting function. Table \ref{tb1} lists the best-fitting parameters for different proper motions components and magnitude types. In general, the random error of proper motions in one component changes from $\sim4.0$ mas yr$^{-1}$ to $\sim8.0$ mas yr$^{-1}$ along with the magnitude changing from 16.0 to 21.0 mag.

\subsection{The color dependence of the systematic and random errors of the proper motions}
\begin{figure*}
\begin{center}
\includegraphics[width=160mm,height=140mm]{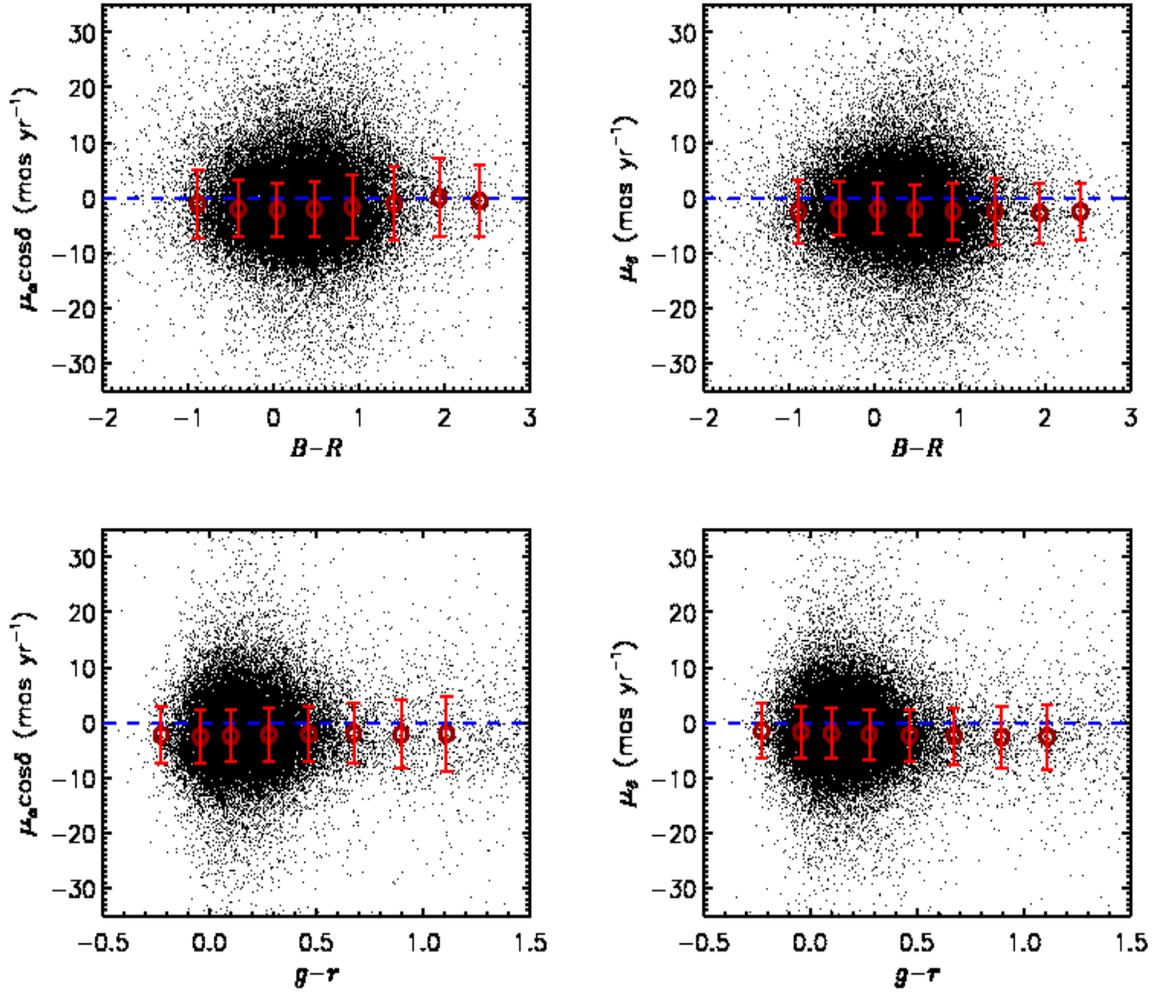}
\caption{Color dependence of the proper motions of quasars in the PPMXL catalog. Top panels: photographic $B-R$ color. Bottom panels: SDSS $g-r$ color. The blue dashed line in each panel shows the proper motions with value of zero. Not all of the quasars in our sample are showed, but the proper motions in each component for all quasars in each color bin are used to derive the systematic and random errors in that color bin, which are showed as red open circles and error bars in each panel.\label{fg7}}
\end{center}
\end{figure*}

Figure \ref{fg7} shows the color dependence of the proper motions of quasars in our sample in each component. The top two panels are based on the photographic $B-R$ color,  and the bottom two panels are based on the SDSS $g-r$ color. The blue dashed line in each panel shows the proper motions with value of zero. In order to derive the dependence of the proper motions in each component on the color, quasars are divided into different color groups with a bin of 0.5 mag for $B-R$ and a bin of 0.2 mag for $g-r$, respectively. For photographic $B-R$ color, quasars with $B-R$ in the range between $-1.25$ and $2.75$ were used.  For SDSS $g-r$ color, quasars with $g-r$ in the range between $-0.4$ and $1.2$ were used. For each color group of quasars, a Gaussian function was used to fit the proper motions in each component in that group, the mean and dispersion of the best-fitting Gaussian function corresponding to the systematic and random errors of proper motions of quasars in that group are showed as red open circle and error bar in each panel of Figure \ref{fg7}.

Figure \ref{fg7} indicate that there is no obvious color dependence of the systematic errors of $\mu_{\alpha}\cos\delta$ and $\mu_{\delta}$ specially in the $g-r$ panel. The rms of the systematic errors in $\mu_{\alpha}\cos\delta$ and $\mu_{\delta}$ for different quasar color groups is $0.8$  and $0.3$ mas yr$^{-1}$ for the photographic $B-R$ color, respectively;  and $0.2$ and $0.4$ mas yr$^{-1}$ for the SDSS $g-r$ color, respectively. In general, the dispersion of system errors of proper motions in one component for quasars in different color group  is smaller than the absolute systematic deviation of the proper motions in that component for all quasars in the sample by a factor of 2 -- 4.

Figure \ref{fg7} indicates that there is no obvious color dependence of the random errors of proper motions  in each component both for $B-R$ and $g-r$. But for quasars with very bluer or redder color, the random errors of proper motions of them are bigger than those for quasars with color in the range of $-0.2 < g-r < 0.7$ . The rms of  the random errors of the proper motions for quasars in different color groups is $0.8$ mas yr$^{-1}$ in $\mu_{\alpha}\cos\delta$ and $0.5$ mas yr$^{-1}$ in $\mu_{\delta}$, respectively. This rms is much smaller than the random error of the proper motions in each component for all quasars in our sample by a factor of 5 -- 10.

\subsection{The position dependence of the systematic and random errors of the proper motions}
\begin{figure*}
\begin{center}
\includegraphics[width=160 mm,height=140 mm]{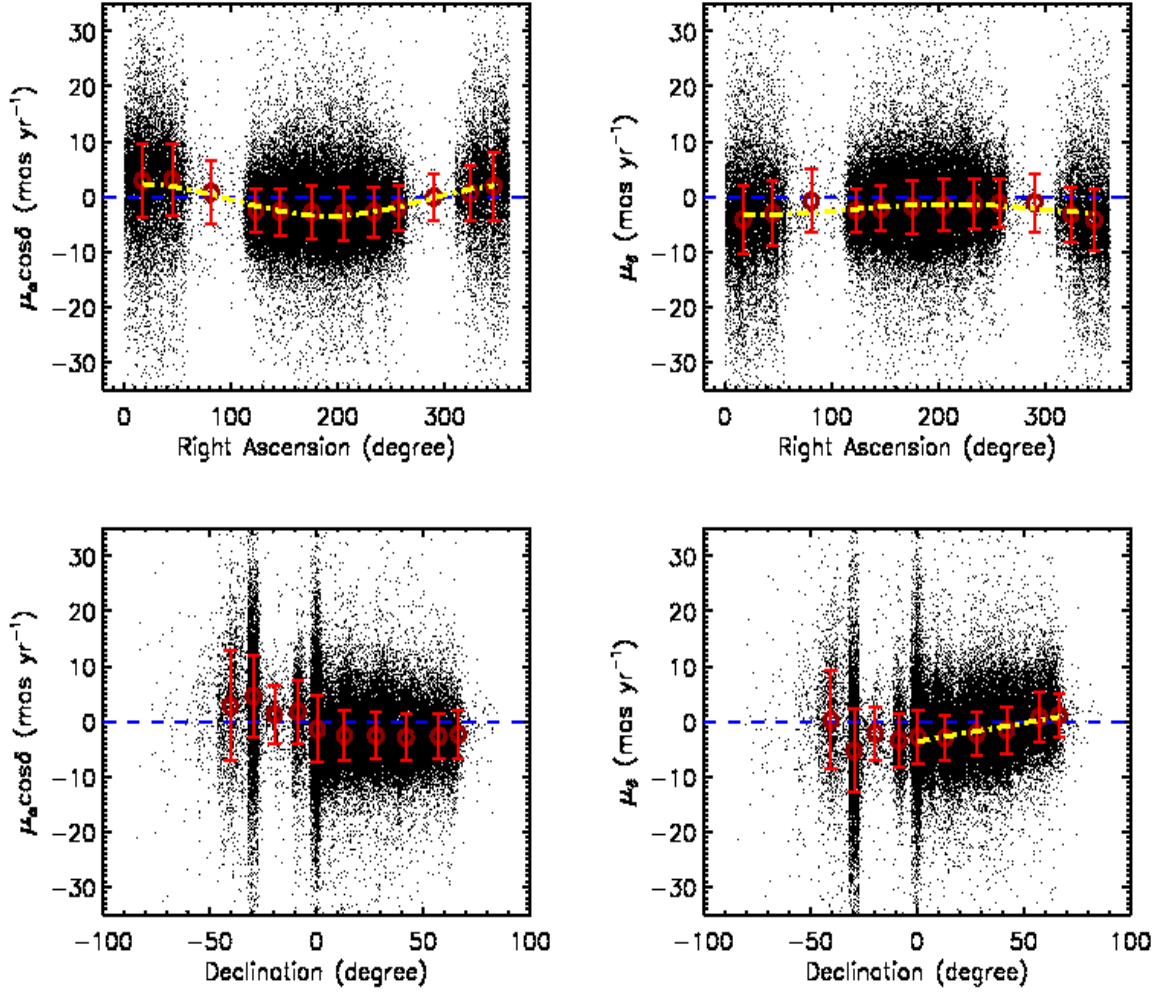}
\caption{Proper motions of quasars in the PPMXL catalog as a function of $\alpha$ and $\delta$. The blue dashed line in each panel shows the proper motions with value of zero. Not all of the quasars in our sample are showed, but the proper motions in each component for all quasars in each position bin are used to derive the systematic and random errors in that position bin, which are showed as red open circles and error bars in each panel. The yellow dash dot lines in the top panels show the best-fitting function: $\overline{\mu}=a+b\times\sin(\alpha-c)$, where $\overline{\mu}$ is the systematic error of proper motions in each component of quasars in each $\alpha$ bin, a, b, and c are the unknown parameters to be fitted, and $\alpha$ is in degree. The yellow dash dot line in the bottom-right panel is the best-fitting line for quasars with $\delta >0$: $\overline{\mu_{\delta}}=-3.65+0.07\delta$, where $\overline{\mu_{\delta}}$ is the systematic error of $\mu_{\delta}$ of quasars in each $\delta$ bin, $\delta$ is in degree.\label{fg8}}
\end{center}
\end{figure*}

Figure \ref{fg8} shows the proper motions of quasars in our sample in each component as a function of right ascension $\alpha$ and declination $\delta$ . The blue dashed line in each panel shows the proper motions with value of zero. In order to derive the dependence of the proper motions in each component on the position, quasars were divided into different position groups with a bin of $35\degr$ in $\alpha$ and a bin of $15\degr$ in $\delta$, respectively. For each position group of quasars, a Gaussian function was used to fit the proper motions in each component in that group, the mean and dispersion of the best-fitting Gaussian function corresponding to the systematic and random errors of proper motions of quasars in that group are showed as red open circle and error bar in each panel of Figure \ref{fg8}.

The top two panels of Figure \ref{fg8} shows  the distribution of $\mu_{\alpha}\cos\delta$ and $\mu_{\delta}$ for quasars in our sample along with $\alpha$. These  panels of Figure \ref{fg8} indicate that, both for $\mu_{\alpha}\cos\delta$ and $\mu_{\delta}$, there are systematic dependence of the systematic errors of the proper motions on $\alpha$. A function of $\overline{\mu}=a+b\times\sin(\alpha-c)$ was used to fit the $\alpha$ dependence of the proper motions in each component, where $\overline{\mu}$ is the systematic error of the proper motions in each component for quasars in each $\alpha$ bin, a, b, and c are the unknown parameters to be fitted, and $\alpha$ is in degree. For $\mu_{\alpha}\cos\delta$, the best-fitting result of the function is  $\overline{\mu_{\alpha}\cos\delta}=-0.59+2.95\sin(\alpha+78.48)$. For $\mu_{\delta}$, the best-fitting result of the function is $\overline{\mu_{\delta}}=-2.30-0.99\sin(\alpha+61.02)$. The best-fitting function for the proper motions in each component is plotted as yellow dash dot line in each top panel of Figure \ref{fg8}. The top two panels of Figure \ref{fg8} also indicate that there is no systematic dependence of the random error of the proper motions in each component on $\alpha$. The dispersion of the random errors in $\mu_{\alpha}\cos\delta$ and $\mu_{\delta}$ for quasars in different $\alpha$ bins is $1.0$ and $0.8$ mas yr$^{-1}$, respectively. This dispersion is smaller than the random error of proper motions for all quasars in our sample by a factor of 4 -- 6.

The bottom two panels of Figure \ref{fg8} shows the distribution of $\mu_{\alpha}\cos\delta$ and $\mu_{\delta}$ along with $\delta$. The bottom-left panel of Figure \ref{fg8} indicates that, for quasars with $\delta >0$, the systematic errors of $\mu_{\alpha}\cos\delta$ in the six $\delta$ bins are smaller than zero. On the other hand, for quasars with $\delta <0$, the systematic errors of $\mu_{\alpha}\cos\delta$ in the four $\delta$ bins are bigger than zero. For quasars with $\delta >0$, there is no $\delta$ dependence of the systematic and random errors in $\mu_{\alpha}\cos\delta$, the dispersion of the systematic and random errors of $\mu_{\alpha}\cos\delta$ for quasars in different $\delta$ bins is 0.5 and 0.7 mas yr$^{-1}$, respectively. For quasars with $\delta <0$, the maximum systematic error of $\mu_{\alpha}\cos\delta=4.6$ mas yr$^{-1}$ is in the nearby of $\delta=-30\degr$. 

The bottom-right panel of Figure \ref{fg8} indicates that, for quasars with $\delta <0$, there is no $\delta$ dependence of the systematic and random errors of $\mu_{\delta}$, but the quasars in the nearby of $\delta=-30\degr$ have the maximum systematic error of $\overline{\mu_{\delta}}=-5.3$ mas yr$^{-1}$. For quasars with $\delta >0$, the bottom-right panel of Figure \ref{fg8} indicates that there is $\delta$ dependence of the systematic error of $\mu_{\delta}$. A line of $\overline{\mu_{\delta}}=-3.65+0.07\delta$ can best fit the $\delta$ dependence of $\mu_{\delta}$, where $\overline{\mu_{\delta}}$ is the systematic error of $\mu_{\delta}$ for quasars in different $\delta$ bins, $\delta$ is in degree. This line is plotted as yellow dash dot line in the bottom-right panel of Figure \ref{fg8}. For quasars with $\delta >0$, there is no $\delta$ dependence of the random errors of $\mu_{\delta}$, the dispersion of the random errors of $\mu_{\delta}$ for quasars in different $\delta$ bins is 0.3 mas yr$^{-1}$.

\subsection{The importance of the 2MASS observation data}
\begin{figure*}
\begin{center}
\includegraphics[width=150 mm,height=180 mm]{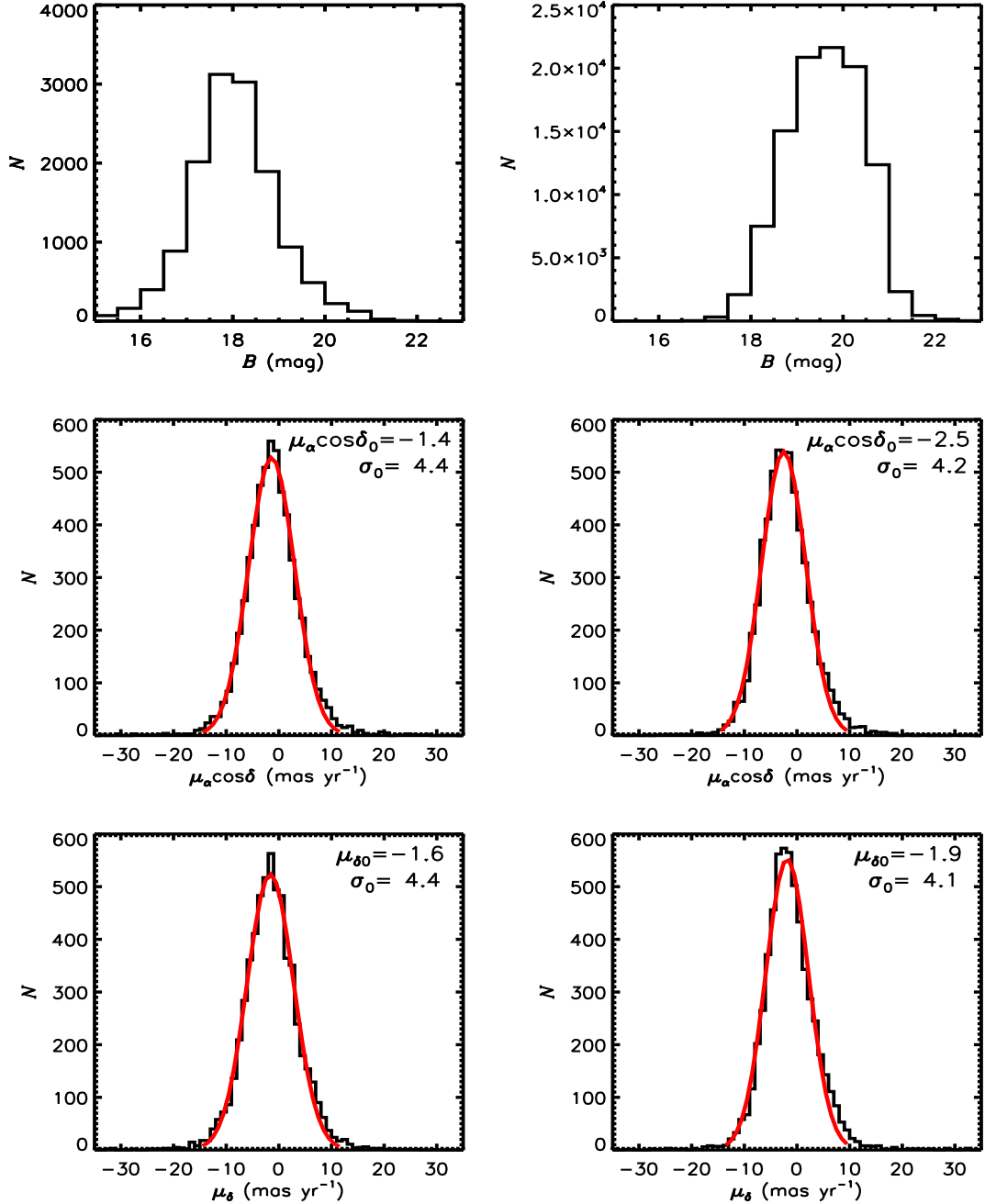}
\caption{Histograms of the photographic $B$ magnitudes and the proper motions of quasars with and without 2MASS observation data in the PPMXL catalog. Left panels are for quasars with 2MASS data, and right panels are for quasars without 2MASS data. Only quasars with $B$ in the range between 17.3 and 18.3 mag are used in the middle and bottom panels showing the proper motions distributions.\label{fg9}}
\end{center}
\end{figure*}

In the previous sections, we have pointed out that there is only 13520 or 12\% quasars in our sample with the 2MASS observation data.  Thus, the proper motions for most of quasars in our sample were derived only from the photographic data. But it is important to compare the derived proper motions with and without the 2MASS data to understand the importance of the 2MASS data in derivation of the proper motions in the PPMXL catalog. A appropriate quasar sample was chosen before this comparison. The top two panels of Figure \ref{fg9} show the photographic $B$ magnitude distributions of quasars with (left panel) and without (right panel) the 2MASS data. These two panels indicate that the $B$ magnitude distributions for these two groups of quasars are very different.  Most of quasars with 2MASS data  are bright and the median of the $B$ magnitude distribution is $\sim18.0$ mag. On the other hand, most of quasars without 2MASS data are faint and the median of the $B$ magnitude distribution is $\sim20.0$ mag. Furthermore, for quasars with $B>19.0$, the number of quasars without 2MASS data is much bigger than that of quasars with 2MASS data, and all quasars without 2MASS data whose $B$ magnitudes are bigger than 17.0 mag. Thus, quasars with $B$ magnitudes in the range between 17.3 and 18.3 mag were chosen as the comparison sample. The number of quasars with and without the 2MASS data in the comparison sample is 6013 and 5878, respectively.

The middle and bottom panels of Figure \ref{fg9} show the histograms of the proper motions in each component for quasars with (left panels) and without (right panels) the 2MASS data in the comparison sample  , respectively. Similar to Figure \ref{fg5}, A Gaussian function was used to fit the  proper motions in each component and was plotted as red solid line in each panel of Figure \ref{fg9}. These four panels in Figure \ref{fg9} indicate that there is no obvious difference in the random errors of the proper motions in both components for quasars with and without 2MASS data. But the systematic error of $\mu_{\alpha}\cos\delta$ for quasars with 2MASS data is much smaller than that for quasars without 2MASS data. The systematic error of  $\mu_{\delta}$ for quasars with 2MASS data is also a little smaller than that for quasars without 2MASS data. Figure \ref{fg9} indicates that the 2MASS data are very important in the derivation of the proper motions in the PPMXL catalog. They can significantly reduce the systematic errors of the derived proper motions in the PPMXL catalog.

\section{Comparison with the USNO-SDSS proper motion}
The USNO-SDSS proper motion uses the USNO-B catalog as the early echo data, but links its zero-point to the galaxies. The SDSS photometric data as the later echo data used in deriving the  USNO-SDSS proper motion are much deeper than the 2MASS photometry used by the PPMXL catalog.  83\%  quasars in the PPMXL catalog are also found in the SDSS DR7.  Thus, it is important to compare the proper motions of quasars in the PPMXL catalog to that in the SDSS catalog to see whether the new SDSS photometric data used in the derivation of the USNO-SDSS proper motion can improve the derived proper motions for the given quasar sample.

Similar to Figure \ref{fg5}, Figure \ref{fg10} shows the histograms of the USNO-SDSS proper motions of quasars cross-identified in the PPMXL and SDSS DR7 catalogs with a bin of 1.0 mas yr$^{-1}$. A Gaussian function was used to fit the USNO-SDSS proper motions in each component and was plotted as red solid line in each panel of Figure \ref{fg10}. Figure \ref{fg10} indicates that the systematic error of the USNO-SDSS proper motions of quasars in the SDSS DR7 catalog is $0.9$ and $0.5$ mas yr$^{-1}$ in $\mu_{\alpha}\cos\delta$ and $\mu_{\delta}$, respectively.  The random error  of the USNO-SDSS proper motions of quasars in the SDSS DR7 catalog is $3.3$ and $3.3$ mas yr$^{-1}$ in $\mu_{\alpha}\cos\delta$ and $\mu_{\delta}$, respectively. The derived systematic and random errors are much smaller than those of the proper motions of quasars in the PPMXL catalog, which indicates the improvement in the derivation of proper motions including the SDSS astrometric and photometric data.

\begin{figure*}
\begin{center}
\includegraphics[width=160mm,height=70mm]{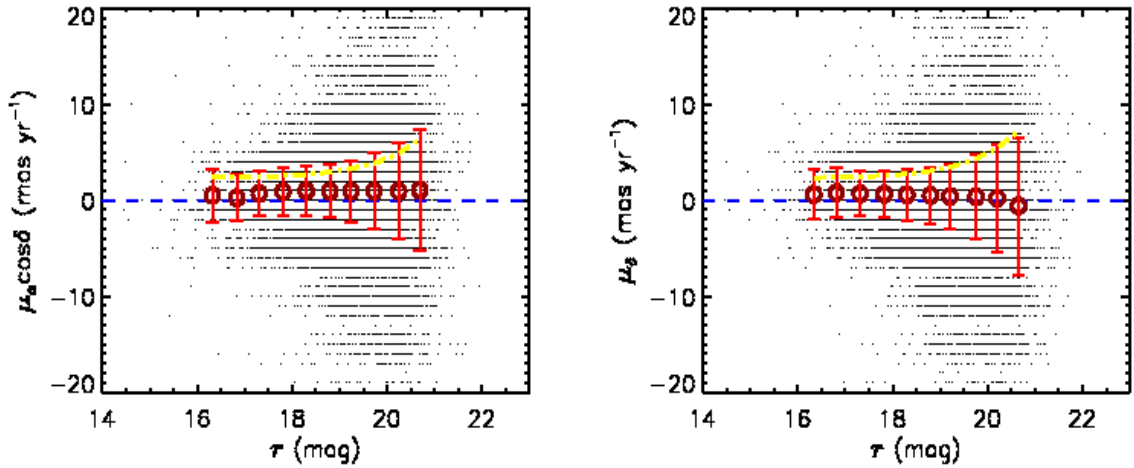}
\caption{Similar to Figure \ref{fg6}, SDSS $r$ magnitude dependence of the USNO-SDSS proper motions for quasars identified  in the PPMXL and SDSS DR7 catalogs.\label{fg11}}
\end{center}
\end{figure*}

Similar to Figure \ref{fg6}, Figure \ref{fg11} shows the SDSS $r$ magnitude dependence of the USNO-SDSS proper motions  in each component of quasars cross-identified in the PPMXL and SDSS DR7 catalogs. Quasars with SDSS $r$ magnitude in the range between 16.0 and 21.0 mag were divided into different magnitude groups with a bin of 0.5 mag. For each magnitude group of quasars, a Gaussian function was used to fit the proper motions in each component in that group. Figure \ref{fg11} indicates that there is no obvious magnitude dependence of the systematic errors of the USNO-SDSS proper motions in the both two components. The rms of the systematic errors of the USNO-SDSS proper motions for quasars in different  magnitude groups is $0.3$  and $0.4$ mas yr$^{-1}$ in $\mu_{\alpha}\cos\delta$ and $\mu_{\delta}$, respectively. This rms is a little smaller than that for quasars in different magnitude groups in the PPMXL catalog.  Figure \ref{fg11} also indicates that the random errors of the USNO-SDSS proper motions in each component increase along with the SDSS $r$ magnitude of quasars.  Similar to Figure \ref{fg6},  the yellow dash dot line in each panel of Figure \ref{fg11} shows the best-fitting function which  describing the SDSS $r$ magnitude dependence of the random error of the USNO-SDSS proper motions in each component. The best-fitting function for the random errors in $\mu_{\alpha}\cos\delta$ is $\sigma_{\mu_{\alpha}\cos\delta}=2.3+3.3\exp^{(r-20.5)}$ mas yr$^{-1}$, and the   best-fitting function for the random errors in $\mu_{\delta}$ is $\sigma_{\mu_{\delta}}=2.3+3.9\exp^{(r-20.4)}$ mas yr$^{-1}$. In general, the random error of the USNO-SDSS proper motions in one component changes from $\sim2.5$ mas yr$^{-1}$ to $\sim6$ mas yr$^{-1}$ along with the SDSS $r$ magnitude changing from 16.0 to 21.0 mag.

\begin{figure*}
\begin{center}
\includegraphics[width=160mm,height=70mm]{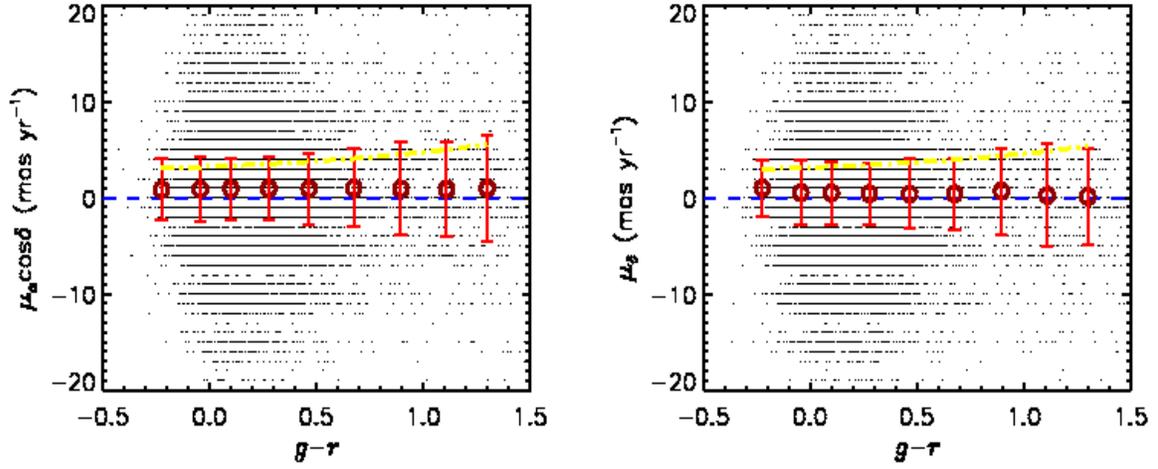}
\caption{Similar to Figure \ref{fg7}, SDSS $g-r$ color dependence of the USNO-SDSS proper motions for quasars identified  in the PPMXL and SDSS DR7 catalogs. The yellow dash dot line in each panel is the best-fitting function which describing the color dependence of the random errors of the USNO-SDSS proper motions: $\sigma_{\mu}=a+b\times\exp^{(m-c)}$, where $\sigma_{\mu}$ is the random error of proper motions in one component, $a$, $b$, and $c$ are the unknown parameters to be fitted, and $m$ is the SDSS $g-r$ color.\label{fg12}}
\end{center}
\end{figure*}

Similar to Figure \ref{fg7}, Figure \ref{fg12} shows the SDSS $g-r$ color dependence of the USNO-SDSS proper motions in each component of quasars  cross-identified in the PPMXL and SDSS DR7 catalogs. Quasars with SDSS $g-r$ color in the range between$-0.4$ and 1.4 were divided into different color groups with a bin of 0.2 mag. For each color group of quasars, a Gaussian function was used to fit the proper motions in each component in that group. Figure \ref{fg12} indicates that there is no obvious color dependence of the systematic errors of the USNO-SDSS proper motions in the both two components. The rms of the systematic errors of the USNO-SDSS proper motions for different quasar color groups is $0.1$  and $0.3$ mas yr$^{-1}$ in $\mu_{\alpha}\cos\delta$ and $\mu_{\delta}$, respectively. This rms is a little smaller than that for quasars in different color groups in the PPMXL catalog.  Figure \ref{fg12} also indicates that the random errors of the USNO-SDSS proper motions in both two components increase along with the SDSS $g-r$ colors of quasars. The color dependence of the random errors of the USNO-SDSS proper motions in each component was fitted and showed as yellow dash dot line in each panel of Figure \ref{fg12} using the similar function used in Figures \ref{fg6} and \ref{fg11}: $\sigma_{\mu}=a+b\times\exp^{(m-c)}$, here $m$ is the SDSS $g-r$ color. The best-fitting function for the random errors in $\mu_{\alpha}\cos\delta$ is $\sigma_{\mu_{\alpha}\cos\delta}=2.4+3.9\exp^{(m-1.5)}$ mas yr$^{-1}$, and the best-fitting function for the random errors in $\mu_{\delta}$ is $\sigma_{\mu_{\delta}}=2.3+3.7\exp^{(m-1.5)}$ mas yr$^{-1}$. In general, the random error of the USNO-SDSS proper motions in one component changes from $\sim2.5$ mas yr$^{-1}$ to $\sim6$ mas yr$^{-1}$ along with the SDSS $g-r$ color changing from $-0.5$ to 1.5.

\begin{figure*}
\begin{center}
\includegraphics[width=160mm,height=140mm]{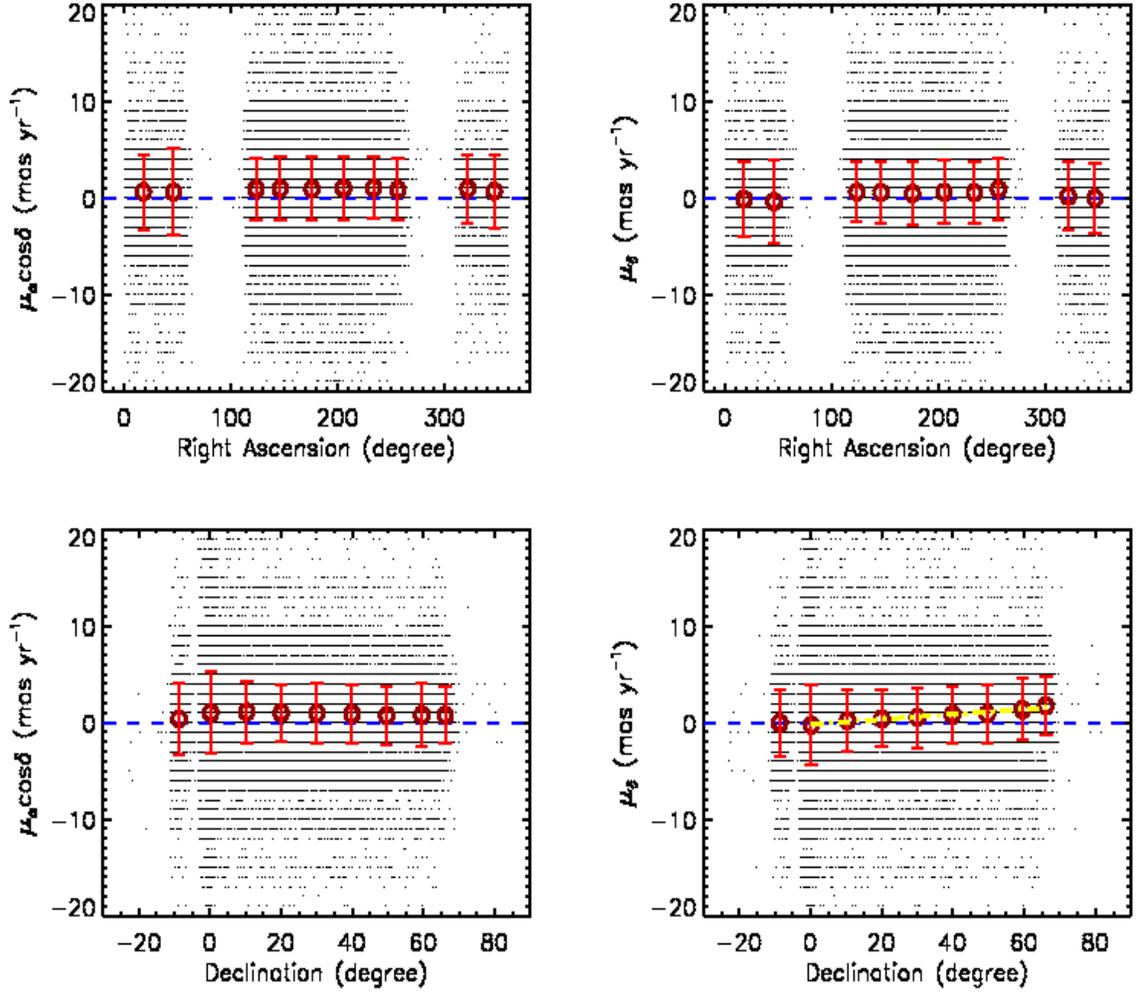}
\caption{Similar to Figure \ref{fg8}, but for the USNO-SDSS proper motions for quasars identified  in the PPMXL and SDSS DR7 catalogs.\label{fg13}}
\end{center}
\end{figure*}

Similar to Figure \ref{fg8}, Figure \ref{fg13} shows the USNO-SDSS proper motions in each component of quasars  cross-identified in the PPMXL and SDSS DR7 catalogs as a function of $\alpha$ and $\delta$. The top two panels of Figure \ref{fg13} indicate that there is no $\alpha$ dependence of the systematic and random errors of the USNO-SDSS proper motions of quasars in the both two components. The rms of the systematic errors of the USNO-SDSS proper motions for  quasars in different  $\alpha$ groups is $0.2$  and $0.4$ mas yr$^{-1}$ in $\mu_{\alpha}\cos\delta$ and $\mu_{\delta}$ , respectively. The rms of the random errors of the USNO-SDSS proper motions for quasars in different $\alpha$ groups is $0.4$  and $0.4$ mas yr$^{-1}$ in $\mu_{\alpha}\cos\delta$ and $\mu_{\delta}$, respectively. The bottom-left panel of Figure \ref{fg13} indicates that there is no obvious $\delta$ dependence of the systematic and random errors of the USNO-SDSS proper motions in $\mu_{\alpha}\cos\delta$. The rms of the systematic and random errors of the USNO-SDSS proper motions in $\mu_{\alpha}\cos\delta$ for quasars in different $\delta$ groups is $0.2$  and $0.4$ mas yr$^{-1}$, respectively. The bottom-right panel of Figure \ref{fg13} indicates that there is no obvious $\delta$ dependence of the random errors of the USNO-SDSS proper motions in $\mu_{\delta}$. The rms of the random errors of the USNO-SDSS proper motions  in $\mu_{\delta}$ for quasars in different $\delta$ groups is $0.4$ mas yr$^{-1}$. Similar to the bottom-right panel of Figure \ref{fg8},  the bottom-right panel of Figure \ref{fg13} indicates the $\delta$ dependence of the systematic errors of the USNO-SDSS proper motions in $\mu_{\delta}$.  For quasars with $\delta >0$,  a line of $\overline{\mu_{\delta}}=-0.18+0.03\delta$ was used to best fit the $\delta$ dependence of  the systematic errors of $\mu_{\delta}$ and showed as yellow dash dot line in the bottom-right panel of Figure \ref{fg13}. The coefficient 0.03 is half of that derived for the $\delta$ dependence of the systematic errors of  $\mu_{\delta}$ for quasars in the PPMXL catalog.

\section{Summaries}
As the very distant extragalactic source, the proper motion of a quasar has been assumed to be zero.  Proper motions of quasars with very high accuracy are needed to check this assumption. In order to detect the secular aberration drift of the extragalactic radio source proper motions caused by the rotation of the Solar System barycenter around the Galactic center, \citet{ti11} derived the absolute proper motions of 555 extragalactic radio sources based on the observation by the very long baseline interferometry (VLBI) between 1990 and 2010. For the most observed radio sources, the inflated position errors are $40\,\mu$as in each coordinate \citep{ti11}. The most of extragalactic radio sources used by \citet{ti11} are quasars, thus, the proper motions derived by \citet{ti11}  are those with the highest accuracy for quasars at present. 

\begin{figure*}
\begin{center}
\includegraphics[width=160mm,height=70mm]{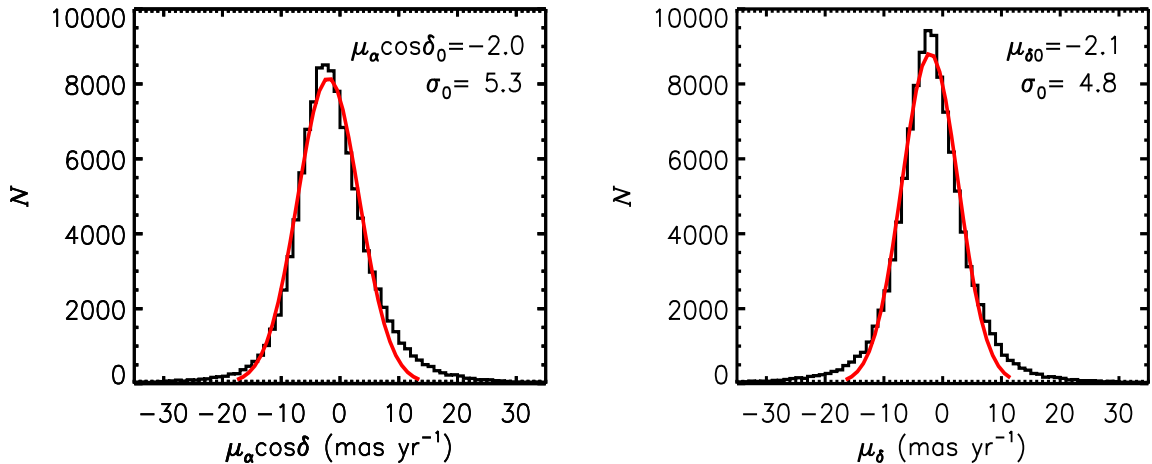}
\caption{Histograms showing the proper motion distributions of quasars in the PPMXL catalog. The red solid line in each panel is the best-fitting Gaussian function. The best-fitting parameters of the Gaussian function: the mean and the dispersion are also labeled in each panel. Only quasars with the proper motions in each component with the absolute value less than 35 mas yr$^{-1}$ are showed in this figure, but the proper motions for all quasars in our sample in each component are used to fit the Gaussian function. The region under the red line in each panel includes proper motion data within the $3\sigma$ from the derived mean.\label{fg5}}
\end{center}
\end{figure*}

\begin{figure*}
\begin{center}
\includegraphics[width=160mm,height=70mm]{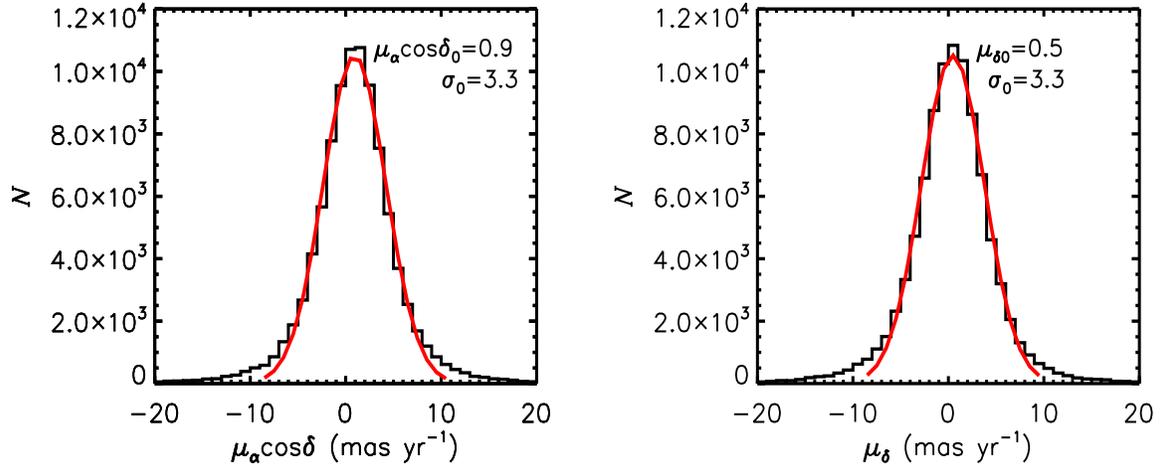}
\caption{Similar to Figure \ref{fg5},  histograms of the USNO-SDSS proper motions for quasars identified  in the PPMXL and SDSS DR7 catalogs. \label{fg10}}
\end{center}
\end{figure*}

\begin{figure*}
\begin{center}
\includegraphics[width=160mm,height=70mm]{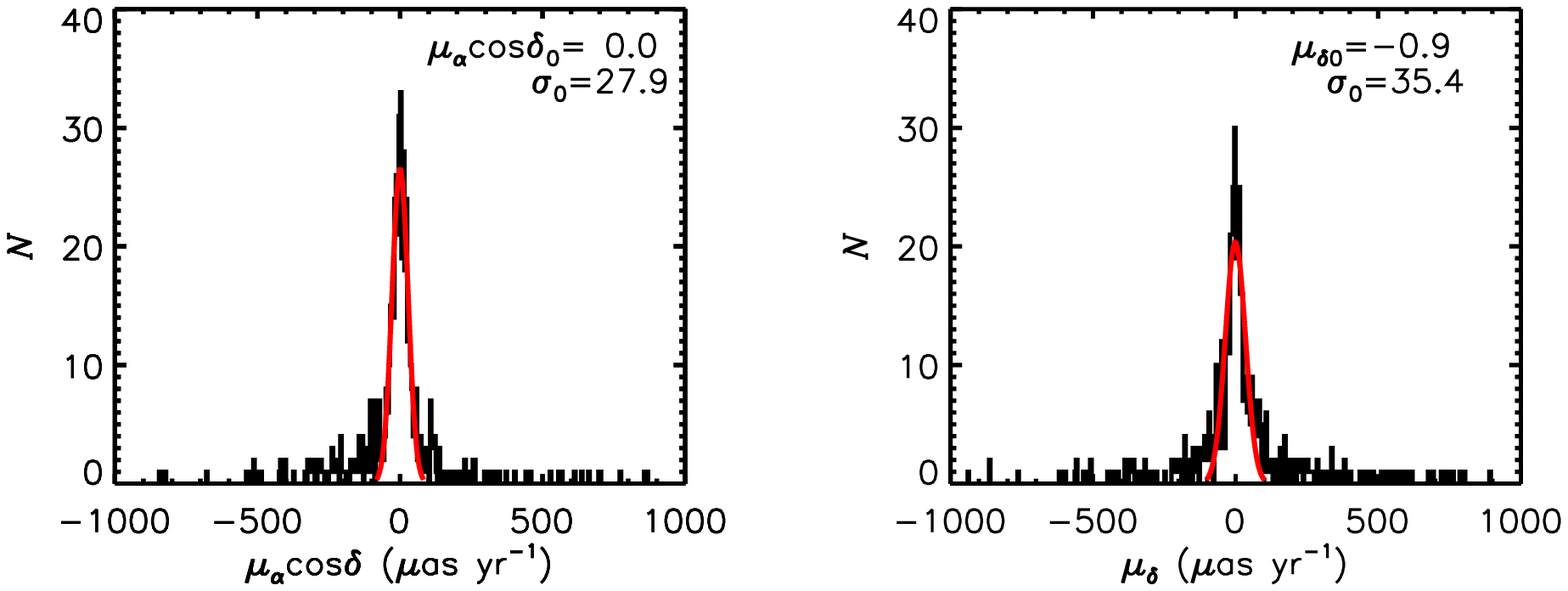}
\caption{Similar to Figure \ref{fg5}, histograms of the proper motions of extragalactic sources derived with the observation by VLBI between 1990 and 2010 \citep{ti11}. Not all 555 sources are showed in this figure, the number of outliers is 13 and 7 in the  $\mu_{\alpha}\cos\delta$ and the $\mu_{\delta}$ panel, respectively.\label{fg14}}
\end{center}
\end{figure*}

Figure \ref{fg14} shows the histograms of the proper motions derived by \citet{ti11} in $\mu_{\alpha}\cos\delta$ and $\mu_{\delta}$ with a bin of $5\,\mu$as yr$^{-1}$. Not all of 555 sources used by \citet{ti11} are showed in Figure \ref{fg14},  and sources with the absolute value of proper motion in each component bigger than 1 mas yr$^{-1}$ are not showed in Figure \ref{fg14}. The number of sources not showed in Figure \ref{fg14} is 13 and 7 in the  $\mu_{\alpha}\cos\delta$ and the $\mu_{\delta}$ panel, respectively. Similar to Figure \ref{fg5}, a Gaussian function was used to fit the proper motions in each component for all 555 sources in the sample of \citet{ti11} and was plotted as red solid line in each panel of Figure \ref{fg14}. Figure \ref{fg14} indicates that the mean of proper motions derived by \citet{ti11} for 555 sources is $0.0$ and $-0.9$ $\mu$as yr$^{-1}$ in  $\mu_{\alpha}\cos\delta$ and $\mu_{\delta}$, respectively. The rms of the derived proper motions by \citet{ti11} for 555 sources is $27.9$ and $35.4$ $\mu$as yr$^{-1}$ in  $\mu_{\alpha}\cos\delta$ and $\mu_{\delta}$, respectively. It should be noted that the unit in Figure \ref{fg14} is $\mu$as yr$^{-1}$ not mas yr$^{-1}$ used in Figures \ref{fg5} and \ref{fg10}. 70\% sources in the sample of \cite{ti11} have proper motions with values in the range within the $3\sigma$ of the mean in each component. 

Figure \ref{fg14} indicates that the mean of proper motions with high accuracy for quasars is close to zero.  Based on a sample of 62 radio sources, \cite{mo11} pointed out that  the proper motions of the radio sources are caused by the changes in their brightness distribution structure. The systematic  motion of quasars can be detected at present is the secular aberration drift with a value of $6.4\pm 1.5$ $\mu$as yr$^{-1}$ \citep{ti11}. In general, the mean of proper motions of quasars can be considered as zero with present observation accuracy. The systematic motion of $6.4$ $\mu$as yr$^{-1}$ of quasars cannot be detected by the present ground-based observation in the  optical band.

Based on the assumption that the mean of proper motions for a sample of quasars should be zero, any deviation from zero of the mean of the proper motions of the sample can be considered as the systematic error of the proper motions, and the dispersion of the proper motions from the mean can be considered as the random error of the proper motions of the sample. A sample of 117053 quasars identified in the PPMXL catalog were used to derive the systematic and random errors in the proper motions of the PPMXL catalog. The systematic error of the proper motions for the whole quasar sample is derived as $-2.0$ and $-2.1$ mas yr$^{-1}$ in  $\mu_{\alpha}\cos\delta$ and $\mu_{\delta}$, respectively. The random error of the proper motions for the whole quasar sample is  $5.3$ and $4.8$ mas yr$^{-1}$ in  $\mu_{\alpha}\cos\delta$ and $\mu_{\delta}$, respectively. 

No obvious magnitude dependence of the systematic errors of the  proper motions in both components are found. The rms of the systematic error of the proper motions in each component for quasars with different magnitudes is $\sim 0.5$ mas yr$^{-1}$.  There is obvious magnitude dependence of the random errors of the proper motions in both components. The random error of the  proper motions in each component can change from  $\sim4.0$ to $\sim8.0$ mas yr$^{-1}$ along with the increase of the magnitude of quasars. No obvious color dependence of the systematic  and random errors of the proper motions in both components are found. The rms of the systematic errors of the proper motions in each component for quasars with different colors is $\sim 0.5$ mas yr$^{-1}$. The rms of the random error of the proper motions in each component for quasars with different  colors is $\sim 0.8$ mas yr$^{-1}$ and $\sim 0.5$ mas yr$^{-1}$ for photographic $B-R$ and SDSS $g-r$, respectively.  

The systematic errors of the proper motions in both components $\overline{\mu}$ are correlated with  $\alpha$ of quasars: $\overline{\mu}\propto \sin\alpha$. No obvious $\alpha$ dependence of the random errors of the proper motions in both components are found. The rms of the random error of the proper motions in each component for quasars with different $\alpha$ is $\sim 1.0$ mas yr$^{-1}$.  For quasars with $\delta<0$, their random errors of the proper motions in both components are bigger than those for quasars with $\delta >0$; and their systematic errors of the proper motions in  both components are very different from those for quasars with $\delta>0$. Especially,  quasars in the neighborhood of $\delta=-30\degr$ have the maximum of the systematic error of  the proper motions in both components $\sim 5$ mas yr$^{-1}$. For quasars with $\delta>0$, there is no obvious $\delta$ dependence of the systematic  and random errors of $\mu_{\alpha}\cos\delta$. The rms of the systematic and random errors of $\mu_{\alpha}\cos\delta$ for quasars with $\delta>0$ is $\sim 0.5$  and $\sim 0.7$ mas yr$^{-1}$, respectively. But for quasars with $\delta>0$, the systematic errors of $\mu_{\delta}$ increase with the $\delta$ and can be best described as $\overline{\mu_{\delta}}\propto 0.07 \delta$. There is no obvious $\delta$ dependence of the random errors of $\mu_{\delta}$ for quasars with $\delta>0$. The rms of the random errors of $\mu_{\delta}$ for quasars with $\delta>0$ is $\sim 0.3$ mas yr$^{-1}$.

The 2MASS observation data are important in the derivation of the proper motions in the PPMXL catalog. The systematic error of $\mu_{\alpha}\cos\delta$ for quasars with 2MASS data is about half of that for quasars without 2MASS data. For comparison, the systematic and random errors of the USNO-SDSS proper motions for a quasar sample identified in the PPMXL and SDSS DR7 catalogs are derived. The systematic error of the USNO-SDSS proper motions is $0.9$ and $0.5$ mas yr$^{-1}$ in $\mu_{\alpha}\cos\delta$ and $\mu_{\delta}$, respectively. The random error of the USNO-SDSS proper motions is $3.3$ and $3.3$ mas yr$^{-1}$ in $\mu_{\alpha}\cos\delta$ and $\mu_{\delta}$, respectively. 

No obvious magnitude dependence of the systematic errors of the  USNO-SDSS proper motions in both components are found. The rms of the systematic errors of the USNO-SDSS proper motions in each component for quasars with different magnitudes is $\sim 0.4$ mas yr$^{-1}$. The random error of the USNO-SDSS  proper motions in each component increases from  $\sim2.5$ mas yr$^{-1}$ to $\sim6.0$ mas yr$^{-1}$ along with the increase of the SDSS $r$ magnitude of quasars. No obvious color dependence of the systematic errors of the  USNO-SDSS proper motions in both components is found. The rms of the systematic errors of the USNO-SDSS proper motions in each component for quasars with different $g-r$ color is $\sim 0.3$ mas yr$^{-1}$. But, the random errors of the USNO-SDSS proper motions in each component can change from $\sim2.5$ mas yr$^{-1}$ to $\sim6.0$ mas yr$^{-1}$ along with the SDSS $g-r$ color of between $-0.5$ and $1.5$.

There is no $\alpha$ dependence of the systematic and random errors of the USNO-SDSS proper motions in both two components.  The rms of the systematic and random errors of the USNO-SDSS proper motions in each component for quasars with different $\alpha$ is $\sim 0.4$ and $\sim 0.4$ mas yr$^{-1}$, respectively. There is no $\delta$ dependence of the random errors of the USNO-SDSS proper motions in both two components.  The rms of the random errors of the USNO-SDSS proper motions in each component for quasars with different $\delta$ is $\sim 0.4$ mas yr$^{-1}$. No obvious $\delta$ dependence of the systematic errors of the  USNO-SDSS proper motions in $\mu_{\alpha}\cos\delta$ is found. The rms of the systematic errors of the USNO-SDSS proper motions in $\mu_{\alpha}\cos\delta$  for quasars with different $\delta$ is $\sim 0.2$ mas yr$^{-1}$.  The systematic errors of the USNO-SDSS proper motions in  $\mu_{\delta}$ increase with the $\delta$ and can be best described as $\overline{\mu_{\delta}}\propto 0.03\delta$.

In general, comparing with the proper motions in the PPMXL catalog, using the SDSS photometric and astrometric data can significantly reduce the systematic and random errors of the derived proper motions. Thus, combining the photographic data and the present observation data obtained from the ground-based telescopes with CCD,  the  proper motions with systematic error less than  1 mas yr$^{-1}$ and random error less than 5 mas yr$^{-1}$ for objects with $r < 20$ mag can be derived.

\begin{deluxetable}{ccccc}
\tablewidth{0pt}
\tablecaption{The best-fitting parameters of function: $\sigma_{\mu}=a+b\times\exp^{(m-c)}$ describing the magnitude dependence of the random errors of proper motions in each component derived from the quasars in the PPMXL catalog. $a$, $b$, and $c$ are unknown parameters to be fitted.\label{tb1}}
\tablehead{\colhead{type of $\sigma_{\mu}$}&\colhead{type of $m$}&\colhead{$a$}&\colhead{$b$}&\colhead{$c$}}
\startdata
$\mu_{\alpha}\cos\delta$&$B$&4.4&3.4&21.0\\
$\mu_{\alpha}\cos\delta$&$r$&3.8&3.3&20.5\\
$\mu_{\delta}$&$B$&4.2&2.1&21.0\\
$\mu_{\delta}$&$r$&3.7&2.6&20.5\\
\enddata
\end{deluxetable}

\acknowledgments
This work has been supported in part by the National Natural Science Foundation of China, No.11073032, 10873016, and 10803007. Z.-Y. W. is supported by the Young Researcher Grant of National Astronomical Observatories, Chinese Academy of Sciences. This research has made use of the VizieR catalogue access tool, operated at CDS, Strasbourg, France.


\begin{thebibliography}{}
%\bibitem[()]{}
\bibitem[Abazajian, et al.(2009)]{ab09} Abazajian, K. N., et al. 2009, \apjs, 182, 543
\bibitem[Bond et al.(2010)]{bo10} Bond, N. A., et al. 2010, \apj, 716, 1
\bibitem[ESA(1997)]{es97} ESA, 1997, The Hipparcos and Tycho Catalogues, ESA, SP-1200 
\bibitem[Fedorov et al.(2010)]{fe10} Fedorov, P. N., Akhmetov, V. S., Bobylev, V. V., \& Bajkova, A. T. 2010, \mnras, 406, 1734
\bibitem[Hambly et al.(2001)]{ha01} Hambly, N. C., Davenhall, A. C., Irwin, M. J., \& MacGillivray, H. T. 2001, \mnras, 326, 1315
\bibitem[H{\o}g et al.(2000)]{ho00} H{\o}g, E., et al. 2000, \aap, 355, L27
\bibitem[Kharchenko(2001)]{kh01} Kharchenko, N. V. 2001, Kinematics and Physics of Celestial Bodies, 17, 409 
\bibitem[Lasker et al.(2008)]{la08} Lasker, B. M., et al. 2008, \aj, 136, 735 
\bibitem[Monet et al.(1998)]{mo98} Monet, D. G., et al. 1998, USNO-A2.0 Catalogue (Flagstaff: US Naval Obs.) 
\bibitem[Monet et al.(2003)]{mo03} Monet, D. G., et al. 2003, \aj, 125, 984 
\bibitem[Mo\'{o}r et al.(2011)]{mo11} Mo\'{o}r, A., Frey, S., Lambert, S. B., Titov, O. A., \& Bakos, J. 2011, \aj, 141, 178
\bibitem[Munn et al.(2004)]{mu04} Munn, J. A., et al. 2004, \aj, 127, 3034 
\bibitem[Munn et al.(2008)]{mu08} Munn, J. A., et al. 2008, \aj, 136, 895
\bibitem[Roeser et al.(2010)]{ro10} Roeser, S., Demleitner, M., \& Schilbach, E. 2010, \aj, 139, 2440
\bibitem[R\"{o}ser et al.(2008)]{ro08} R\"{o}ser, S., Schilbach, E., Schwan, H., Kharchenko, N. V., Piskunov, A. E., \& Scholz, R.-D. 2008, \aap, 488, 401
\bibitem[Schneider et al.(2010)]{sc10} Schneider, D. P., et al. 2010, \aj, 139, 2360
\bibitem[Skrutskie et al.(2006)]{sk06} Skrutskie, M. F., et al. 2006, \aj, 131, 1163 
\bibitem[Souchay et al.(2009)]{so09} Souchay, J., et al. 2009, \aap, 494, 799
\bibitem[Titov et al.(2011)]{ti11} Titov, O., Lambert, S. B., \& Gontier, A. -M. 2011, \aap, 529, A91
\bibitem[Urban et al.(2000)]{ur00} Urban, S. E., Wycoff, G. L., Makarov, V. V. 2000, \aj, 120, 501
\bibitem[V\'{e}ron-Cetty \& V\'{e}ron(2010)]{ve10} V\'{e}ron-Cetty, M. -P., \& V\'{e}ron, P. 2010, \aap, 518, A10
\bibitem[Wu et al.(2009)]{wu09} Wu, Z. Y., Zhou, X., Ma, J., \& Du, C. H. 2009, \mnras, 399, 2146
\bibitem[Wu et al.(2011)]{wu11} Wu, Z. Y., Zhou, X., Ma, J., \& Du, C. H. 2011, \aj, 141, 104
\bibitem[Zacharias et al.(2004)]{za04} Zacharias, N., Urban, S. E., Zacharias, M. I., Wycoff, G. L., Hall, D. M., Monet, D. G., \& Rafferty, T. J. 2004, \aj, 127, 3043
\bibitem[Zacharias et al.(2010)]{za10} Zacharias, N., et al. 2010, \aj, 139, 2184
\end{thebibliography}
\end{document}